\documentclass{icrc2009}

\usepackage{graphicx}   % for including figures
\usepackage{caption}    % for captions
\usepackage{fixltx2e}
\usepackage{stfloats}
\fnbelowfloat
\usepackage{url}

\newcommand{\shorttitle}[1]%
{\markboth{Proceedings of the 31\MakeLowercase{$^{st}$} ICRC, {\L}\'{o}d\'{z}
2009}{#1} }
\begin{document}
\title{Simulation of the Cosmic Ray Moon Shadow in the Geomagnetic field}

\author{\IEEEauthorblockN{Giuseppe Di Sciascio\IEEEauthorrefmark{1} and
                           Roberto Iuppa\IEEEauthorrefmark{1}\IEEEauthorrefmark{2}\\
}
\\
\IEEEauthorblockA{\IEEEauthorrefmark{1} INFN, Sezione Roma Tor
Vergata, Via della Ricerca Scientifica 1, Rome - Italy}
 \IEEEauthorblockA{\IEEEauthorrefmark{2} Dipartimento di Fisica,
 Universit\'a Roma Tor Vergata, Rome - Italy}}

% please write the preseter's name and short title (3-4 words maximum)
%    which will appear at the header of the even pages.
\shorttitle{G. Di Sciascio and R. Iuppa Moon Shadow Simulation}
\maketitle

\begin{abstract}
An accurate MonteCarlo simulation of the deficit of primary cosmic
rays in the direction of the Moon has been developed to interpret
the observations reported in the TeV energy region until now. 
Primary particles are propagated trough the
geomagnetic field in the Earth-Moon system. The algorithm is
described and the contributions of the detector resolution 
and of the geomagnetic field are disentangled.
\end{abstract}
\section{Introduction}
Since the galactic cosmic rays are hampered by the Moon, a deficit
of cosmic rays in its direction is expected (the so-called "Moon
shadow"). The Moon shadow is an important tool to calibrate the
performance of an air shower array. In fact, the size of the
deficit allows a measurement of the angular resolution and the
position of the deficit allows the evaluation of the absolute
pointing accuracy of the detector. In addition, charged
particles are deflected by the geomagnetic field
by an angle inversely proportional to their energy. 
The observation 
of such a displacement provides a direct check of the relation 
between shower size and primary energy, thus calibrating the detector.

A detailed Monte Carlo simulation of cosmic ray propagation in the
Earth-Moon system is mandatory to understand the Moon
shadow phenomenology and to compare the observed westward
displacement with the expectations in order to disentangle the
geomagnetic effect from some possible experimental biases.
\section{Monte Carlo Simulation}
\subsection{Simulation strategy}
When the emission of photons by a given gamma-ray source has to be
simulated, a classical approach can be followed: once its energy
spectrum is known, it is enough to decide for how long the source
emission has to be reproduced to calculate the number of expected
events which have to be sampled. Of course, if the response of an
Earth-based array has to be simulated, for each one of these
primary particles also the induced shower and the detector
response have to be calculated. The easiest way to do that, is to
shoot up the particles from the detector\footnote{Actually, from a
suitably-choosen area around it.} to the source. This trick allows
to concentrate the efforts of the simulation only on the showers
which effectively construct the signal.

Nonetheless, this method cannot be used for the Moon-shadow as it
is, because of two main reasons.
\begin{enumerate}
 \item The Moon shadow is not a signal, but a "lack" in the background.
\item the effect is provoked by charged particles and not by
photons. This implies that we must take into account the effect of
the electro-magnetic fields on their trajectories.
\end{enumerate}

The first argument is the more relevant one. In fact, if we wanted
to remain faithful to the approach of simulating only the
particles which reach the detector, we should compute the showers
of the background sorrounding the Moon and then take off the ones
coming exactly from within Moon. In such terms, the simulation is
likely unfeasible, at least for low energy threshold experiments. 
To be precise,if the detector energy threshold is a few hundreds of GeV, 
even by considering a square sky window sorrounding the Moon not so large (e.g. 
$10^{\circ}$), too many showers should have to be simulated to reproduce 
even just one year of data taking. 

It is evident that the Moon shadow requires a different 
strategy of simulation. It is better to treat the Moon like a  standard source
and then reverse the
amplitude of the signal. In the end, both the gamma-source case
and the Moon shadow case reduce to a perturbation of the
cosmic-rays background. The only difference is the sign of the
perturbation.

Because of their electric charge the cosmic rays do not proceed
straight from the Moon to the Earth, but their trajectories are
bent by the geomagnetic field. Since this effect plays a crucial
role in the final result of the physics analysis, it must not be
underestimated during the simulation. Especially for those
particle which have low energy, the bending effect is very strong
and a realistic prediction is possible only if the simulation
accounts correctly for the intensity and the direction of the
geomagnetic field. What is more, notice that when the particles 
are sent back to the Moon from the detector, the sign of their charge 
must be reversed to properly reproduce the direction of the deviation.
\subsection{The generation and the detection of the showers}
The air showers development in the atmosphere has been generated
with the CORSIKA v. 6.500 code including the QGSJET-II.03 hadronic
interaction model for primary energy above 80 GeV and the FLUKA
code for lower energies \cite{corsika}. Cosmic ray spectra have
been simulated in the energy range from 30 GeV to 1 PeV following
the relative normalization given in \cite{wiebel-sooth}, resulting
from a global fit of main experimental data. About $10^8$ showers
have been sampled in the zenith angle interval 0-60 degrees. The
secondary particles have been propagated down to a cut-off energy
of 1 MeV.

At present, the only experiment able to observe the Moon shadow with 
high statistical significance and energy threshold well below 1 TeV 
is ARGO-YBJ \cite{ICRC09-Moon}. Therefore we reproduced 
an ideal detector placed in YangBaJing (4300 m a.s.l.) having geometrical 
features similar to ARGO-YBJ and a duty-cycle of $90\%$. We used a GEANT4-based code \cite{GEANT4}.
 The trigger threshold 
has been set to $20$ charged particles over the whole detector.
The core positions have been randomly sampled in an energy-dependent area large up to 10$^3$ $\times$ 10$^3$
m$^2$, centered on the detector.
\subsection{The geomagnetic model}
It has been already noticed that if a primary cosmic ray (energy $E$, 
charge $Z$) traversing the geomagnetic field is observed by a detector 
placed in YangbaJing, its trajectory shows a deviation along the 
East-West direction\cite{amenomori} 
\footnote{No deviation is expected along the North-South one.} 
which in first approximation can be written as:
\begin{equation}
\label{eq:DipoleDisplacement}
\Delta\vartheta\simeq-1.58^{\circ}\frac{Z}{E[\textrm{TeV}]}
\end{equation}
The sign is set according to the usual way to represent the East-West
projection of the Moon maps (see Fig. \ref{fig:SimulatedMaps}). The eq. \ref{eq:DipoleDisplacement} can be
easily derived by assuming that the geomagnetic field is provoked
by a pure dipole laying in the centre of the Earth (see Appendix).
Nonetheless this approximation, which is derived for nearly vertical 
primaries, is not enough when the primaries energy is below few TeV.

To perform a numerical estimate of the bending effect, it is necessary 
to adopt a model of the magnetic field in the Earth-Moon system. The 
roughest one is the so-called Virtual Dipole Model (VDM), whose name 
is self-explaining. A better choice is the Tsyganenko-IGRF model 
(T-IGRF hereafter) \cite{Tsyganenko}, which accounts for both internal 
and external magnetospheric sources. We compared the effect on the particle 
trajectories of VDM and T-IGRF, in both cases finding non negligible
differences 
with respect to the $-1.58^{\circ}Z/E[\textrm{TeV}]$ formula. Among the 
two models themselves, we observed discrepancies up to the $\sim 15\%$ 
level ($4^{\circ}\div7^{\circ}$) for sub-TeV primary energies, mainly due to the field intensity 
near the Earth surface. Since the T-IGRF model accounts for more factors, 
we refer to it hereafter.
\begin{figure}[!ht]
 \centering
 \includegraphics[width=3.0in]{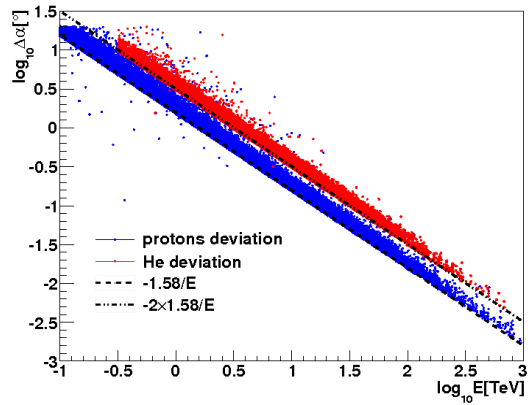}
 \caption{Deviation induced by the geomagnetic field on protons and He 
nuclei. Each point refers to a simulated shower. 
Both the arrival direction and the date of the propagation are randomly 
sampled, respectively from an isotropic distribution within the sky and 
from a uniform distribution over 2008. The analytical trends obtained 
from the equation \ref{eq:DipoleDisplacement} 
are also shown.}
  \label{fig:EWDisplacements}
 \end{figure}
 In Fig. \ref{fig:EWDisplacements} you can 
appreciate the analytical trend (eq. \ref{eq:DipoleDisplacement}) 
together with the actual East-West displacement calculated applying the T-IGRF model 
for protons and He nuclei. The analytical approach clearly underestimates 
the East-West deviation.
\begin{figure}[!ht]
 \centering
 \includegraphics[width=2.5in]{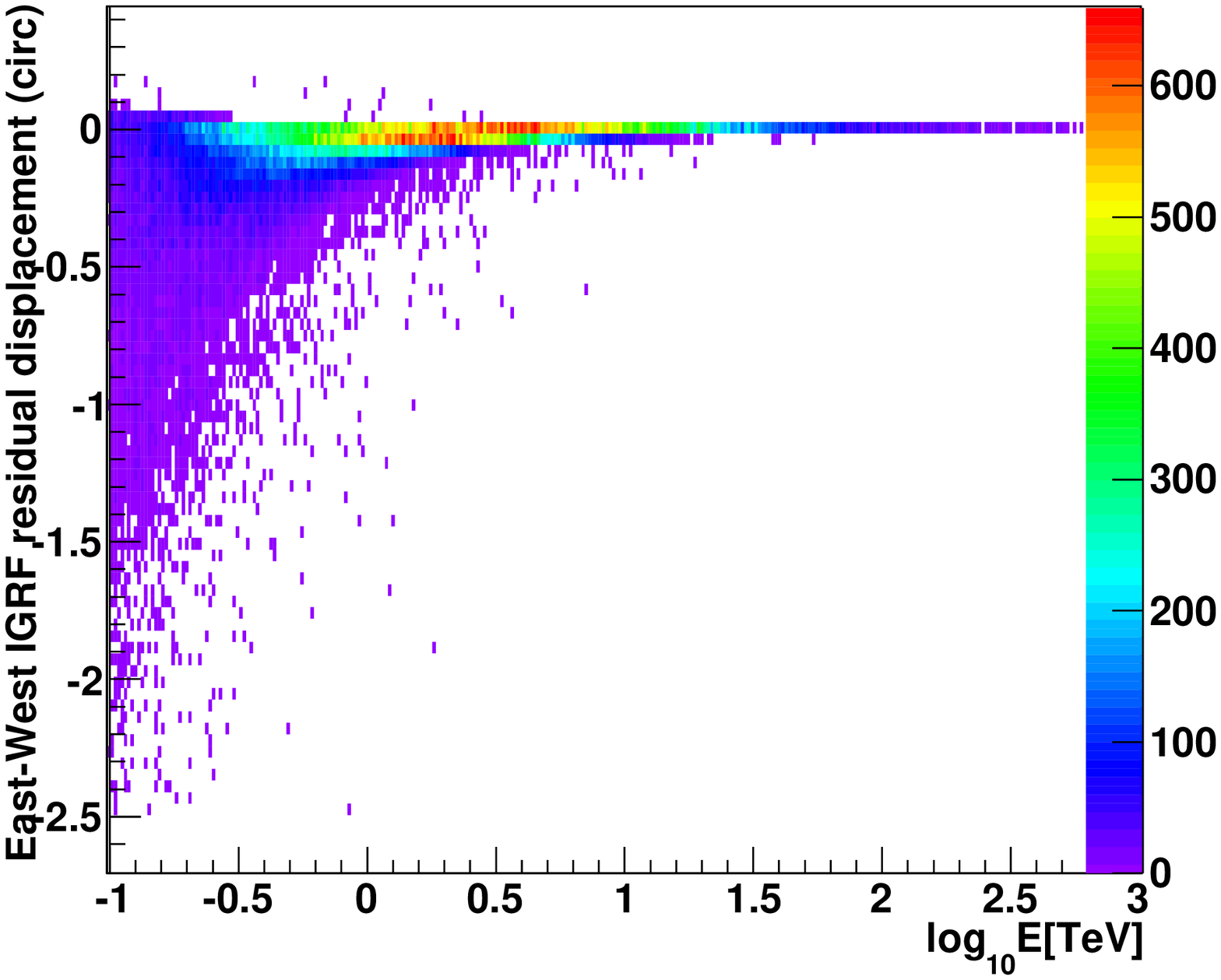}
\includegraphics[width=2.5in]{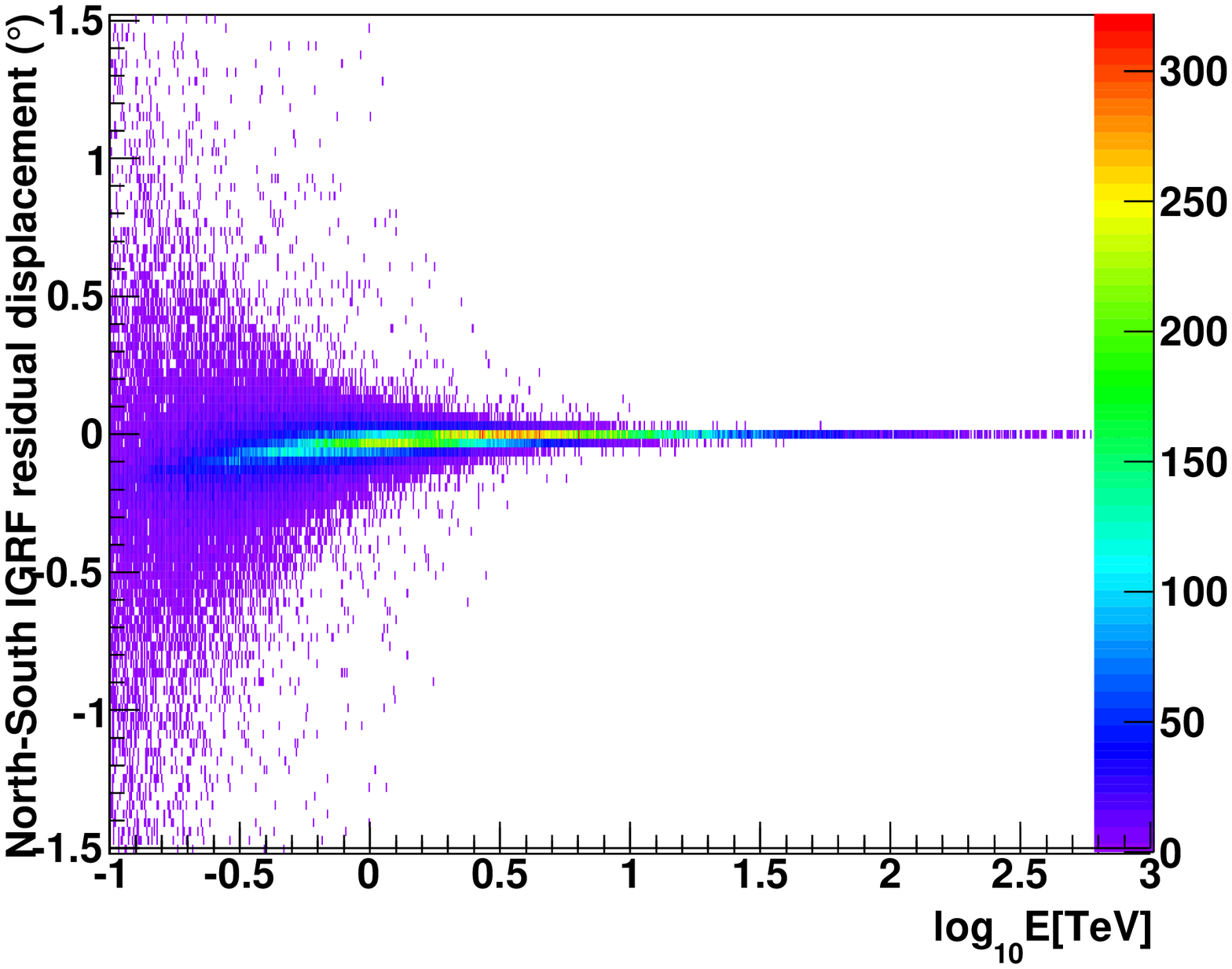}
 \caption{Residual displacement with respect to the analytical expectation. 
The deviation is calculated by applying the the T-IGRF model (see text). The color scale 
represents the number of showers laying on the single pixel.}
  \label{fig:GeomagneticTrend}
 \end{figure}

The Fig. \ref{fig:GeomagneticTrend} shows the differences of the 
T-IGRF-induced deviation with respect to the leading term 
$1.58^{\circ}Z/E\textrm{[TeV]}$. The upper (lower) panel contains such a 
residual deviation along the East-West (North-South) direction as a function 
of the primary energy. Altough there are no effects for energies 
$E>10\textrm{ TeV}$, below few TeV the residual displacement can reach 
$1^\circ$. Notice that unlike the analytical approach would 
suggest, the North-South deviation of the single primary is non-null, 
being zero only on average.
\subsection{Moon shadow simulation}
\begin{figure}[!ht]
  \centering
  \includegraphics[width=0.4\textwidth]{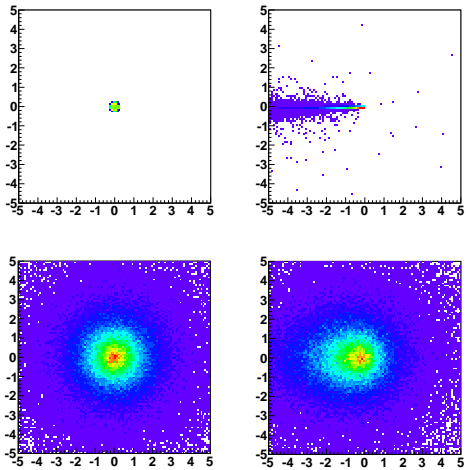}
  \caption{Folding different contributions to the Moon signal. Upper part of the figure: Moon as it would be 
observed by a perfect detector without geomagnetic field; only the magnetic field is switched on. Lower part: only the detector 
is switched on; both the magnetic field and the detector are switched on. 
The maps are drawn using equatorial coordinates (declination VS corrected right ascension). 
Only the showers triggering the detector are shown ($N>20$). 
The color scale represents the intensity of the signal.}
  \label{fig:SimulatedMaps}
 \end{figure}

By following the procedure described before, we can obtain the Moon shadow maps represented in Fig. 
\ref{fig:SimulatedMaps}. 
There can be appreciated the displacement induced by the geomagnetic field. The long tail of the left part of the 
up-right map is made by the lowest energy particles (below 1 TeV) which are more deviated. Concerning the bottom-left 
map, the detector by itself provides the smearing of the signal, leaving intact the circular symmetry, as expected.

It is possible to study the effects of the finite angular resolution of the detector 
and of the geomagnetic field separately.

As already noticed, if we consider the magnetic deviation but not the detector, the circular symmetry 
of the signal is broken only along the East-West direction (see Fig. \ref{fig:SimulatedMaps}, top-right map). 
That make us confident that 
only the smearing due to the angular resolution affects the signal along the north-south direction, 
thus allowing its determination. Actually, we stress that what we determine 
by considering the spread of the signal along the North-South direction, cannot be properly named 
angular resolution, because the Moon is not a point-like source and its own finite angular width (half a degree) 
contributes to the spread. The superposition of the two 
effects can be easily visualized in case of gaussian Point Spread Function (PSF):
\begin{eqnarray}
 RMS=\sigma\sqrt{1+\left(\frac{0.13^{\circ}}{\sigma}\right)^2} \nonumber
\end{eqnarray}
where the root mean square of the signal $RMS$ is related to the variance $\sigma^2$ of the PSF. 
The contribution of the Moon size to the RMS is (not) dominant when $\sigma$ is low (high), 
i.e. at high (low) particle multiplicities. Just to be explicit, the difference between $RMS$ and $\sigma$ is 
$20\%$ if $\sigma=0.2^{\circ}$, less than $5\%$ if $\sigma>0.4^{\circ}$, and only $1.7\%$ if $\sigma=0.7^{\circ}$.
\begin{figure}[!ht]
  \centering
  \includegraphics[width=0.4\textwidth]{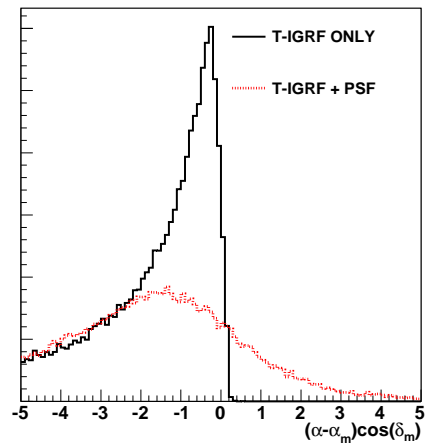}
  \caption{Effect of the PSF along the east-west direction. The continous black line represents the Moon shadow 
deformed by the geomagnetic field as it would appear to an ideal detector. By considering also the effect of the 
detector PSF, the diplacement of the signal peak is enhanced, moreover the well-known smearing effect. The figures 
represent only protons.}
  \label{fig:EffectOfTheAngularResolution}
 \end{figure}
 \begin{figure}[!h]
  \centering
  \includegraphics[width=0.4\textwidth]{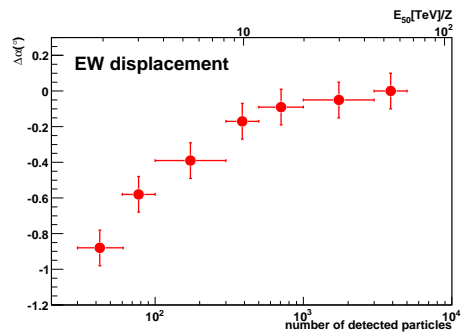}
  \caption{Expected displacement of the Moon shadow in the East-West direction as a function of multiplicity. 
The upper scale refers to the median energy of rigidity (TeV/Z) in each multiplicity bin (shown by the horizontal errors).}
  \label{fig:EWDisplacement}
 \end{figure}

Finally, it is worth to invert the switching order of the detector effect and of the geomagnetic field. 
In Fig. \ref{fig:EffectOfTheAngularResolution} the effect of the angular resolution on the East-West projection 
of the Moon is shown. Because of the signal asymmetry induced by the magnetic field, such an effect provides 
not only the smearing, but also a further displacement of the signal peak. The West tail of the shifted signal, 
in fact, has a larger weight than the sharp East one and tends to pull the signal in its direction.

In Fig. \ref{fig:EWDisplacement} the total eastward displacement is plot versus the number of detected particles.
\section{Conclusions}
We reproduced the Moon shadow effect as observed by an ideal detector located
at high altitude. Much attention has been devoted to the simulation of 
the geomagnetic field bending the particle trajectories. We have been able 
to disentangle the contribution to the final signal of the geomagnetic field 
and of the detector PSF respectively. We quantified the contribution 
of the finite Moon disc to the angular resolution estimation. For a comparison 
with the data collected by the ARGO-YBJ experiment, see \cite{ICRC09-Moon}.

\section*{Appendix}
Here is shown how to obtain the formula
\ref{eq:DipoleDisplacement}.
Since only the magnetic field is supposed to act upon the
particles trajectories, the Lorentz equation can be written as:
\begin{equation}
 \label{eq:IntegralLorentzEquation}
\mathbf x(t)=\mathbf x_0+\mathbf
v_0\,t+\frac{Zec^2}{E}\int_0^t\textrm{d}\tau\,\int_0^\tau\textrm{d}\alpha\,
\frac{\textrm{d}\mathbf x}{\textrm{d} \alpha}\times \mathbf
B(\mathbf x,\alpha)
\end{equation}
where:
\begin{itemize}
 \item $\mathbf x(t)$ is the particle position at the time $t$;
 \item $\mathbf x_0$ and $\mathbf v_0$ are the initial position and velocity of
the particle;
 \item $Ze$ and $E$ are its charge and its (constant) energy;
 \item $\mathbf B(\mathbf x,t)$ is the magnetic field, which in principle can
vary both with respect to the position and to the time.
\end{itemize}

If it is possible to write down an explicit functional form for
$B(\mathbf x,t)$, an attempt to solve the equation
\ref{eq:IntegralLorentzEquation} can be made. On the contrary,
especially when the variation with the time cannot be easily
summarized with an analytical formula, a numerical solution is
unavoidable\footnote{I.e. what has been done in the main part of
this paper.}.

The equation \ref{eq:IntegralLorentzEquation} explicitly shows the
perturbation induced by the magnetic field on the straight
trajectory ($\mathbf x(t) = \mathbf x_0+\mathbf v_0t$). This
suggests an iterative method to determine the solution, which can
be expressed as the series:
\begin{eqnarray}
 \label{eq:SolutionSeries}
\mathbf x(t)= \mathbf x_{\mathcal O(B^0)}(t)+\mathbf x_{\mathcal
O(B^1)}(t)+\dots\nonumber
\end{eqnarray}
where $x_{\mathcal O(B^0)}(t)=\mathbf x_0 +\mathbf v_0t$ is the
unperturbed (straight) trajectory and for the higher orders holds:
\begin{eqnarray*}
\label{eq:SolutionSeriesTerm}
\Delta\mathbf x_{\mathcal O(B^{i+1})}(t)=\\
=\frac{Zec^2}{E}\int_0^t\textrm{d}\tau\int_0^\tau\textrm{d}\alpha
\frac{\textrm{d}\mathbf x_{\mathcal O(B^i)}}{\textrm{d}
\alpha}\times \mathbf B(\mathbf x_{\mathcal O(B^i)},\alpha)\\
\textrm{\ \ \ $i=0,1,\dots$} 
\end{eqnarray*}
where $\Delta\mathbf x_{\mathcal O(B^{i+1})}(t)=\mathbf x_{\mathcal O(B^{i+1})}(t)-(\mathbf x_0+\mathbf v_0\,t)$
is the displacement from the unperturbed trajectory at the time
$t$. Being content with the first-order approximation, it can be
obtained:
\begin{eqnarray*}
\Delta\mathbf x(t)\simeq\frac{Zec^2}{E}\mathbf
v_0\times \int_0^t\textrm{d}\tau\,\int_0^\tau\textrm{d}\alpha\,
\mathbf B(\mathbf x_0+\mathbf v_0\alpha,\alpha)
\end{eqnarray*}
or:
\begin{eqnarray}
 \label{eq:FirstOrderDisplacement}
\Delta\mathbf x(t)\simeq\frac{Z}{E}\mathbf
v_0\times\mathcal I_{\mathbf B}(t;\mathbf x_0, \mathbf v_0)
\end{eqnarray}
where $\mathcal I_{\mathbf B}(t;\mathbf x_0, \mathbf v_0)$ is
the integral of the magnetic field along the straight trajectory,
whose value depends only on the time of the motion ($t$) and on
its initial conditions ($\mathbf x_0$ and $\mathbf v_0$).

Since the phenomenon studied concerns ultra-relativistic particles
and fixing for a moment the initial position and the final time,
the equation \ref{eq:FirstOrderDisplacement} becomes:
\begin{eqnarray}
\Delta\mathbf x\simeq\frac{Z}{E}\ \hat{\mathbf v}_0\times\mathcal
I_{\mathbf B}(\hat{\mathbf v}_0)\nonumber
\end{eqnarray}
In short, on a first approximation the displacement depends only
on the ratio charge-to-energy of the primary and on the initial
direction of its ultrarelativistic motion (versor $\hat{\mathbf
v}_0$).\footnote{The second dependence is not trivial at all.
Because of the difference in the field-integrals, two identical
particles having the same energy and starting together from the
Moon can drift from the straight trajectory differently according
to their different initial directions.}

Let us consider only the lowest order of the geomagnetic field
multipoles-expansion, i.e. the \emph{dipole term}:
\begin{eqnarray}
\mathbf B(\mathbf x)=\frac{3(\mathbf b\cdot\mathbf x)\mathbf
x-x^2\mathbf b}{x^5}\nonumber
\end{eqnarray}
where $\mathbf b$ has intensity $b\approx8.1\cdot10^{27}$ T m$^3$ 
and the south magnetic pole is supposed to have coordinates 
$78.3^{\circ}$ South, $111.0^{\circ}$ East.
By setting $\hat{\mathbf v}_0||\mathbf x_0$ 
(\emph{vertical direction approximation}) and integrating 
from YangBaJing to a distance $\sim60$ Earth's radii, the equation 
\ref{eq:DipoleDisplacement} is immediately obtained.

\end{document}